# Parameterization of nearshore wave breaker index

Chi Zhang [a,b,*], Yuan Li [a,b], Yu Cai [a,b], Jian Shi [a,b], Jinhai Zheng [b,*], Feng Cai [c], Hongshuai Qi [c]

[a] *State Key Laboratory of Hydrology-Water Resources and Hydraulic Engineering, Hohai University, 1 Xikang Road, Nanjing, 210098, China*

[b] *College of Harbor, Coastal and Offshore Engineering, Hohai University, 1 Xikang Road, Nanjing, 210098, China*

[c] *Third Institute of Oceanography, Ministry of Natural Resources, 178 Daxue Road, Xiamen, 361005, China*

*Corresponding author. E-mail address*: zhangchi@hhu.edu.cn (C. Zhang), jhzheng@hhu.edu.cn (J. Zheng).

## Highlights

- The composite dependence of wave breaker index on both the offshore wave steepness and the normalized local water depth.

- A new formula for wave breaker index used for parametric wave transformation modelling leads to improved wave height prediction especially under wave conditions with small offshore wave steepness.

- Two counteractive physical mechanisms to explain the composite relationships.

## Abstract：

The performances of phase-averaged parametric nearshore wave transformation models depend significantly on a reliable estimate of the wave breaker index $\gamma$ (the breaker height-to-depth ratio), a free model parameter that essentially controls the amount of breaking energy dissipation. While the previous studies have suggested separate relationships between $\gamma$ and the offshore wave steepness ($s_0$) or the normalized local water depth ($kh$), the parameterization of $\gamma$ still requires further investigation considering a wider variety of conditions and a sounder physical basis. In this study, we use the field datasets of wave height and the inverse modelling approach to reveal a composite dependence of $\gamma$ on both $s_0$ and $kh$. Specifically, the results show a positive dependence of $\gamma$ on $kh$ for larger $s_0$, and a negative dependence of $\gamma$ on $kh$ for smaller $s_0$. Based on such composite relationships, a new $\gamma$ formula is proposed, and its performance is verified against the available datasets of wave height in three coasts and 14 laboratory tests. Implementation of this new formula in a parametric wave model leads to the error reduction of wave height prediction by 10~24% (mean = 19%) relative to seven widely used models in literatures. In particular, a remarkably higher model accuracy is obtained under wave conditions with small offshore wave steepness, which is important for studying onshore sediment transport and beach recovery. Two counteractive physical mechanisms for wave nonlinearity effects, namely the breaking intensification mechanism and the breaking resistance mechanism, are suggested to explain the opposite $\gamma$-$kh$ relationships within different ranges of $s_0$.

## 1. Introduction

Phase-averaged parametric wave models based on the energy balance equation have been widely used in coastal engineering practice to predict the nearshore wave height transformation of random waves (e.g., Battjes and Janssen, 1978; Thornton and Guza, 1983; Janssen and Battjes, 2007; Zheng et al., 2008; Zhang et al., 2018; Shao et al., 2018; Shi et al., 2019). In shallow waters, wave breaking is the dominant process controlling wave height distribution across the seabed profile, which also affects nonlinear wave shape, wave-induced currents, near-bed sediment transport and beach evolution (e.g., Ruessink et al., 1998, 2001; Zheng et al., 2014; Ma et al., 2017). Therefore, reliable representation of the breaking-induced energy dissipation is one of the key factors affecting the performances of parametric wave models.

Most of the existing parametric random wave transformation models calculate the wave breaking energy dissipation using the breaking probability approach. The models employ a specific probability distribution of random wave heights, and at a given water depth ($h$), estimate the fractional energy loss of the waves that exceed a threshold height. Battjes and Janssen (1978) (hereafter BJ1978) employed a clipped Rayleigh distribution truncated at a maximum wave height, assuming all breaking waves have a same height. Thornton and Guza (1983) (hereafter TG1983) found that wave heights in surf zone corresponded well with the full Rayleigh distribution and proposed a weighted Rayleigh distribution for breaking waves. An alternative weighed function was proposed by van der Westhuysen (2010) to consider wave nonlinearity effects. Baldock et al. (1998) (hereafter B1998) provided a full Rayleigh distribution with a single threshold breaker height ($H_b$) above which all individual waves were assumed to break, leading to an improved model performance in the unsaturated surf zone. The energy dissipation formulation in B1998 was further revisited by Janssen and Battjes (2007) (hereafter JB2007) and Alsina and Baldock (2007) to correct for a shoreline singularity in very shallow water.

Other studies have focused on the breaker index ($\gamma$, the maximum/breaker wave height-to-depth ratio) in the parametric wave models, a free model parameter that essentially controls the amount of breaking energy dissipation at different cross-shore locations. In practice, $\gamma$ is usually calibrated in the model to reproduce the observed root-mean-square wave height ($H_{rms}$) or significant wave height ($H_s$) (Apotsos et al., 2008). Empirical formulas for both cross-shore constant and cross-shore varying $\gamma$ have been derived and incorporated in the parametric wave models, being mainly the functions of the offshore wave steepness ($s_0 = H_0/L_0$, where $H_0$ and $L_0$ are the offshore wave height and wave length), the normalized local water depth ($kh$, where $k$ and $h$ are the local wave number and water depth) or the local bed slope. Based on laboratory datasets of wave height, Battjes and Stive (1985) proposed a hyperbolic tangent relationship between $\gamma$ and $s_0$. Based on the field datasets and using the inverse modelling approach, Ruessink et al. (2003) (hereafter R2003) suggested that $\gamma$ increases linearly with $kh$ and does not depend on the bed slope or $s_0$. Apotsos et al. (2008) tested and calibrated several parametric wave models with field experiments and found that the best-fit $\gamma$ universally increases with the increasing $H_0$. More recently, Salmon et al. (2015) (hereafter S2015) and Lin and Sheng (2017) (hereafter L2017) developed $\gamma$ formulas related to the local bed slope and $kh$. In general, the above studies significantly improved the model accuracy by considering $\gamma$ as a tunable model parameter, while made less efforts to the physical interpretation for the proposed relationships.

It should be noted that the $\gamma$ investigated in the present study is in principle an empirical breaker index (equivalent to $H_b/h$) calibrated in the parametric random wave models. All referred to as a wave height-to-depth ratio, it is by definition different from the field observed $H_{rms}/h$ or $H_s/h$ (e.g., Sallenger and Holman, 1985; Raubenheimer et al., 1996; Sénéchal et al., 2001, 2005; Power et al., 2010), the laboratory observed $H_b/h_b$ of mainly regular waves where $h_b$ is the breaker depth at the breakpoint (e.g., Rattanapitikon and Shibayama, 2000; Goda, 2010; Liu et al., 2011; Robertson et al., 2013, 2017), or the field observed $H_b/h_b$ of individual waves at their incipient breaking (e.g., Robertson et al., 2015a, 2015b). Although the above ratios are not necessarily or directly comparable (Battjes and Stive, 1985; Ruessink et al., 2003), some qualitative inconsistency among them from the physical perspectives might be worth mentioning. For example, several studies proposed $\gamma$ increases with the increasing $kh$ (e.g., Ruessink et al., 2003; Lin and Sheng, 2017), which is opposite to the field observation that $H_{rms}/h$ and $H_s/h$ decreases with the increasing $kh$ (e.g., Raubenheimer et al., 1996; Sénéchal et al., 2001, 2005; Power et al., 2010), as well as to the observed negative dependence of $H_b/h_b$ on $h_b/L_0$ (e.g., Rattanapitikon and Shibayama, 2000; Goda, 2010; Robertson et al., 2015b). van der Westhuysen (2010) argued that the positive dependence of $\gamma$ on $kh$ could be explained by the difference in breaking dissipation intensity related to different degrees of wave nonlinearity. Moreover, several studies suggested $\gamma$ increases with the increasing offshore wave parameters $s_0$ or $H_0$ (e.g., Battjes and Stive, 1985; Apotsos et al., 2008), which seems to contrast with the traditional understanding with the surf similarity parameter (Battjes, 1974) that, the spilling wave breaker more likely occurring with a larger $s_0$ has a smaller breaker index. In general, the parameterization of $\gamma$ still requires further investigations by considering the comprehensive influences of the offshore wave parameters, the local water depth, the breaker characteristics and the surf zone state in order to obtain a sounder physical basis as well as the improved model predictability.

On the basis of the work by Ruessink et al. (2003), the present study uses the inverse modelling approach to revisit the wave breaker index in the parametric wave models with respect to a wider variety of conditions. The composite influences of $s_0$ and $kh$ on $\gamma$ are revealed, and a new $\gamma$ formula is proposed. The performance of this new formula is verified against both field and laboratory datasets and compared with seven widely used models in literatures. The physical relevance for the proposed relationships is also discussed.

## 2. Model formulation

The parametric wave model applied in the present study is the same to that of Ruessink et al. (2003). In principle, it solves the cross-shore random wave transformation with the energy balance equation reads

$$\frac{\partial}{\partial x}(\frac{1}{8}\rho g H_{rms}^2 c_g \cos\theta) = -D_b \tag{1}$$

where $x$ is the coordinate and positive onshore, $\rho$ is the water density, $g$ is the gravity acceleration, $H_{rms}$ is the root-mean-square wave height, $c_g$ is the group velocity, $\theta$ is the wave angle, $D_b$ is the wave breaking-induced energy dissipation rate calculated with the formulation of Baldock et al. (1998):

$$D_b = \frac{\alpha}{4} f_p \rho g \exp[-(\frac{H_b}{H_{rms}})^2](H_b^2 + H_{rms}^2) \tag{2}$$

where $\alpha$ is of order one and controls the level of energy dissipation, $f_p$ is the peak wave frequency, $H_b$ is the maximum/breaker height expressed as (Battjes and Janssen, 1978)

$$H_b = \frac{0.88}{k}\tanh(\gamma h \frac{k}{0.88}) \tag{3}$$

where $k$ is the wave number calculated with the peak frequency, $h$ is the water depth, $\gamma$ is a free model parameter equivalent to the maximum/breaker wave height-to-depth ratio which will be calibrated using the inverse modelling approach in Section 4.

The above model formulations and the unique inverse modelling approach employed in this study follow closely the methodology of Ruessink et al. (2003), with the only difference appearing to be the derived $\gamma$ formula. This allows a straightforward evaluation of the specific effects of the new $\gamma$ formula in Section 5, by directly comparing the present modelling results to those of Ruessink et al. (2003).

## 3. Field datasets used for calibration and verification

Three field datasets of wave height are used for model calibration and verification, covering a wide range of wave conditions over different bed profiles. Here we provide a summary of the datasets, while the detailed descriptions of the field measurements are referred to Ruessink et al. (2003) and Apotsos et al. (2008).

For model calibration, we use the 313-h $H_{rms}$ data collected at 14 cross-shore locations (extending to the water depth of about 4.5 m) during Duck94 experiment (Elgar et al., 1997) from September 21, 1994 to October 4, 1994. Offshore wave data in Duck94 experiment was measured with an array of bottom-mounted pressure sensors in 8-m water depth (Long, 1996). Offshore root-mean-square wave height and peak wave period ranged from 0.12 to 1.98 m and 4.1 to 9.8 s, respectively. The angle of wave incidence ranged from -0.81 to 0.82 rad. The offshore wave steepness can be calculated using the linear wave theory with the measured offshore root-mean-square wave height and the peak wave period.

For model verification, the 529-h data collected at Duck from October 4, 1994 to October 26, 1994 is used. During this time period, offshore root-mean-square wave height, peak wave period and wave angle ranged from 0.29 to 2.88 m, 4.5 to 11.6 s and -0.60 to 0.84 rad, respectively. In addition, the measured wave heights at 6 cross-shore locations (extending to the water depth of about 4 m, spanning 841-h, from October 15, 1998 to November 19, 1998) at Egmond, the Netherlands (Ruessink et al., 2001), and at 5 cross-shore locations (extending to the water depth of about 6 m, spanning 816-h, from May 25, 1994 to June 28, 1994) at Terschelling, the Netherlands (Ruessink et al., 1998; Houwman, 2000) are also used for model verification. Offshore waves at Egmond were recorded by a directional buoy in 15-m water depth, with the root-mean-square wave height, peak wave period and wave angle ranging from 0.30 to 3.90 m, 4.2 to 10.5 s and -1.50 to 1.08 rad, respectively. At Terschelling, the offshore root-mean-square wave height, peak wave period and wave angle ranged from 0.12 to 1.84 m, 3.0 to 12.8 s and -0.52 to 0.75 rad, respectively.

## 4. Calibration of the wave breaker index

Following Ruessink et al. (2003), variation of the breaker index ($\gamma$) is inversely calculated using Eqs. (1), (2) and (3). For a given cross-shore transect of $H_{rms}$, the calculation procedure is summarized in the following steps: (1) fit a smooth curve of $H_{rms}$ using the Hermite interpolation with a grid spacing of 1 m; (2) calculate $c_g$, $\theta$ and $k$ at each grid point using the linear wave theory and Snell's law with the offshore peak period, wave angle and tidal level; (3) calculate wave energy flux at each grid point with $H_{rms}$, $c_g$ and $\theta$; (4) calculate the energy dissipation rate $D_b$ as the gradient of wave energy flux using Eq. (1) at each measurement point; (5) calculate $H_b$ at each measurement point using Eq. (2) with $D_b$, $f_p$ and $H_{rms}$ based on the Newton iteration method; (6) calculate $\gamma$ at each measurement point using Eq. (3) with $H_b$, $k$ and $h$.

It is noted that Ruessink et al. (2003) only retained the $\gamma$ values for $D_b > 15$ N/ms to avoid spurious results. However, we find this treatment is not necessary in the present study as it may cause the miss of some important information. In this study, all $\gamma$ values for $D_b > 0$ N/ms are retained. The influence of such threshold selection is shown in Figure 1. It is found that the data of $D_b > 15$ N/ms accounts for only a small portion (18%) of all data, and it mainly represents the wave conditions when $s_0$ is relatively large (approximately $> 0.02$). Therefore, the calibration of $\gamma$ under wave conditions with smaller offshore wave steepness is likely not included in Ruessink et al. (2003).

Figure 2 presents the relationship between $\gamma$ and $kh$. As shown, the $\gamma$ data for $D_b > 15$ N/ms is in good agreement with the formula ($\gamma = 0.76kh + 0.29$) of Ruessink et al. (2003), which describes a positive correlation between $\gamma$ and $kh$. However, this relationship no longer applies if all $\gamma$ data is considered. Especially, the $\gamma$ data seems to follow a different trend under wave conditions with smaller $s_0$ and $D_b$. Moreover, the $\gamma$ data for $D_b > 15$ N/ms (mainly corresponding to the energetic wave conditions with large $s_0$) is generally higher than other data, implying that the breaker index depends on both the local water depth and the offshore wave parameters.

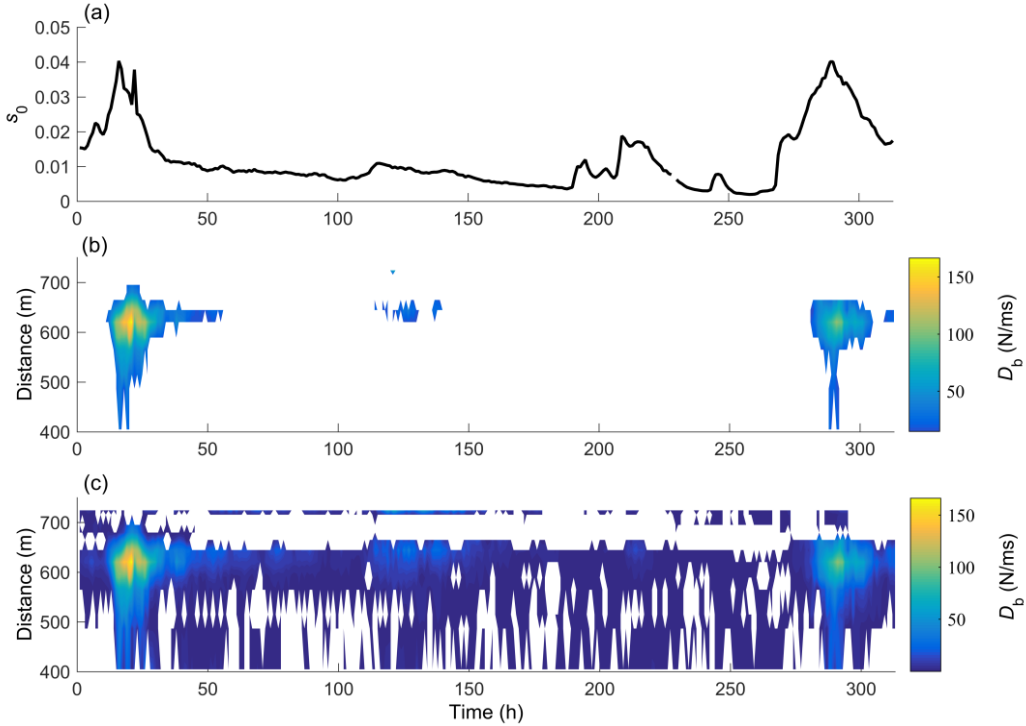

**Fig. 1.** The data at Duck: (a) time series of offshore wave steepness, (b) time-space diagram of the energy dissipation rate $D_b$ retaining the values greater than 15 N/ms, (c) time-space diagram of the energy dissipation rate $D_b$ retaining all positive values. Time = 0 corresponds to September 21, 1994, 12:00 EST. Distance is relative to the offshore measurement location.

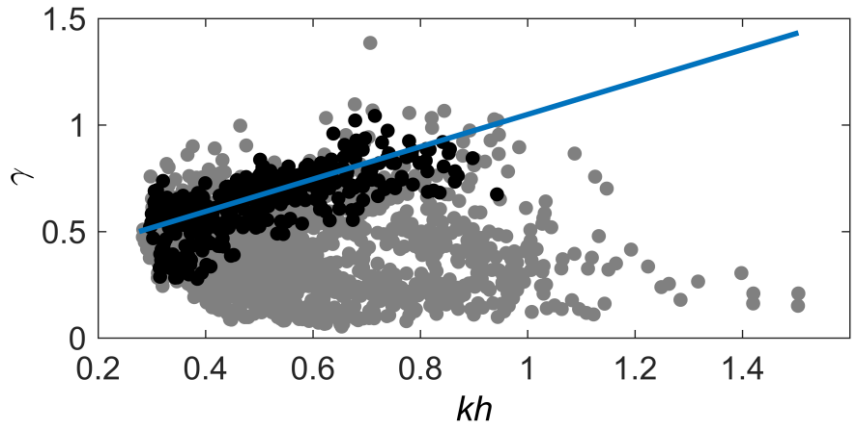

**Fig. 2.** The breaker index $\gamma$ versus $kh$ based on the estimates with $D_b > 15$ N/ms (black points) and all estimates with $D_b > 0$ N/ms (grey points). The solid line is the $\gamma$ formula of Ruessink et al. (2003).

In Figure 3, we plot the $\gamma$-$kh$ relationships with respect to thirteen different groups of $s_0$ values, divided by the value interval of 0.003. It is interesting to find that, when $s_0$ is relatively low (e.g., $s_0 < 0.023$), $\gamma$ generally decreases as $kh$ increases (Figure 3a-3g). In contrast, when $s_0$ is relatively high (e.g., $s_0 > 0.026$), $\gamma$ generally increases as $kh$ increases (Figure 3i-3m). To the authors' knowledge, such opposite $\gamma$-$kh$ relationships corresponding to different ranges of the offshore wave steepness have not been noted before. The physical interpretation for this finding will be discussed in Section 6.

The exponential curves can well describe the data trends in Figure 3, and we propose the $\gamma$ formula with a new form as

$$\gamma = f(s_0)\exp[g(s_0)\times kh] \tag{4}$$

where $f(s_0)$ and $g(s_0)$ are functions of the offshore wave steepness.

To parameterize $f(s_0)$ and $g(s_0)$, the best-fit values based on the data in Figure 3 and Eq. (4) are plotted versus the average $s_0$ of the corresponding groups, as shown in Figure 4. The results show good correlations. The second-order polynomials and the ln function can be used to parameterize $f(s_0)$ and $g(s_0)$ with the highest values of the correlation coefficient ($R^2 = 0.84$ and $R^2 = 0.98$), respectively.

Finally, we propose a new $\gamma$ formula which matches well with the data ($R^2 = 0.86$) (Figure 5) and is expressed as

$$\gamma = (237\,s_0^2 - 34.81\,s_0 + 1.46)\cdot\exp\left[1.96\ln(38.64\,s_0)\times kh\right] \tag{5}$$

It is noted that the applicability of Eq. (5) might be limited by the available data ranges of $s_0$ (0~0.05) and $kh$ (0.3~1.2). For $s_0$ or $kh$ beyond these ranges, the nearest upper or lower limits are thus recommended to be used instead.

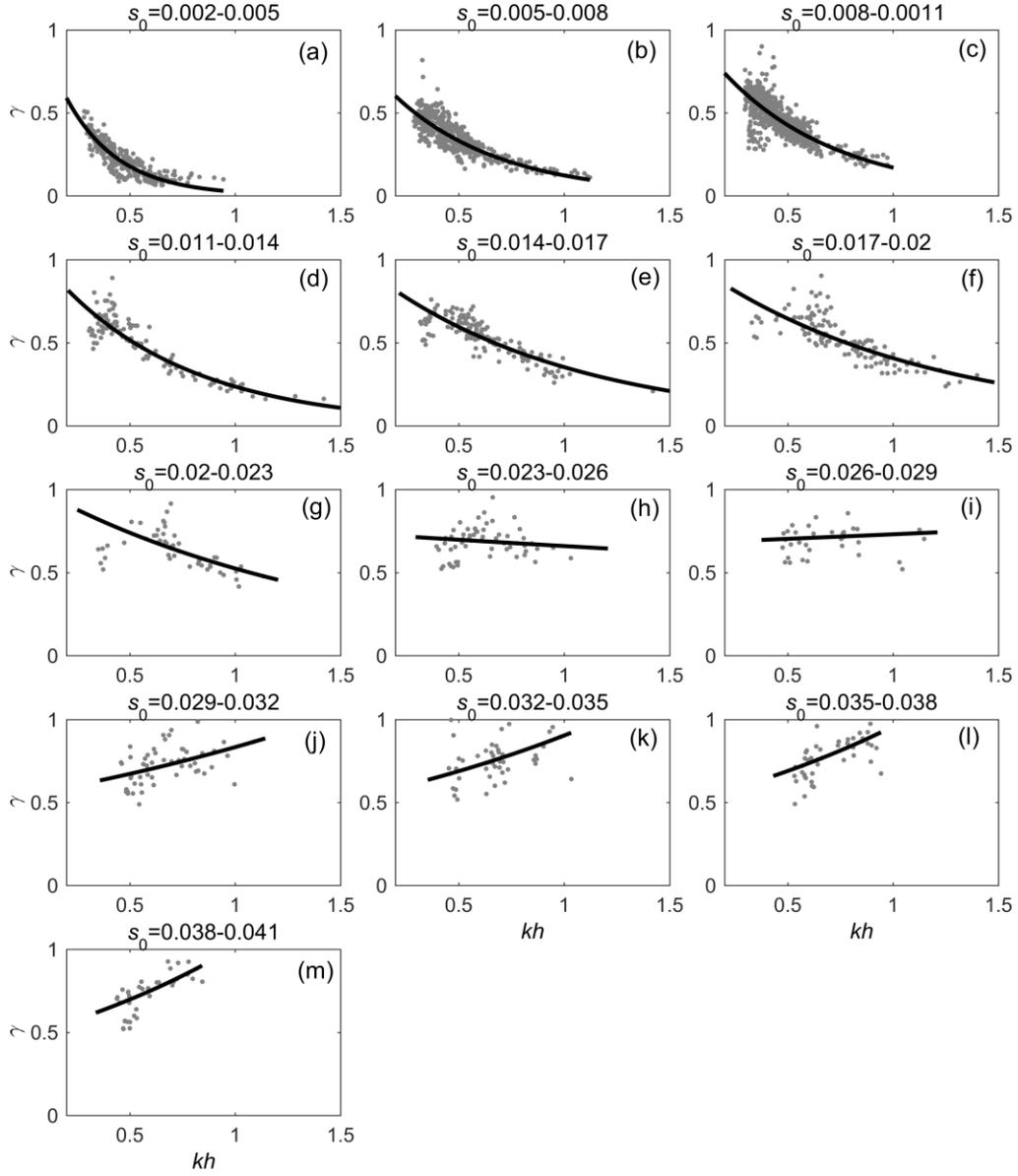

**Fig. 3.** The breaker index $\gamma$ versus $kh$ with respect to different groups of $s_0$ values. The solid lines are exponential curves fitted by least square method.

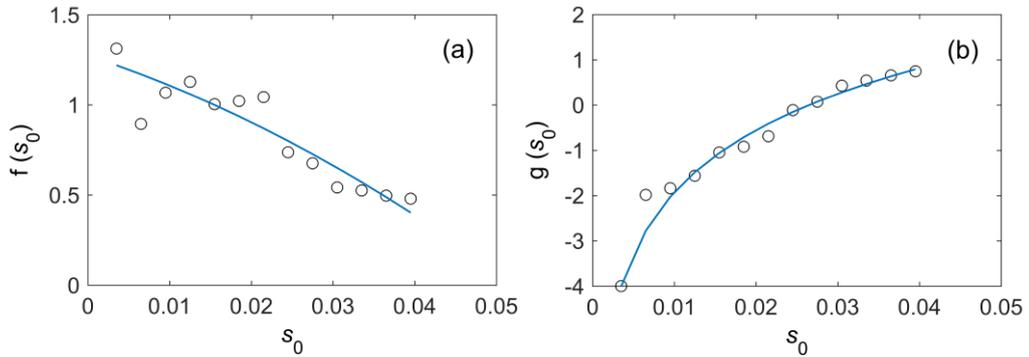

**Fig. 4.** The best-fit values of f($s_0$) (a) and g($s_0$) (b) as functions of $s_0$, based on the data in Figure 3 and Eq. (4). The solid lines are the fitted second-order polynomials (a) and ln function (b) with the correlation coefficients $R^2 = 0.84$ and $R^2 = 0.98$, respectively.

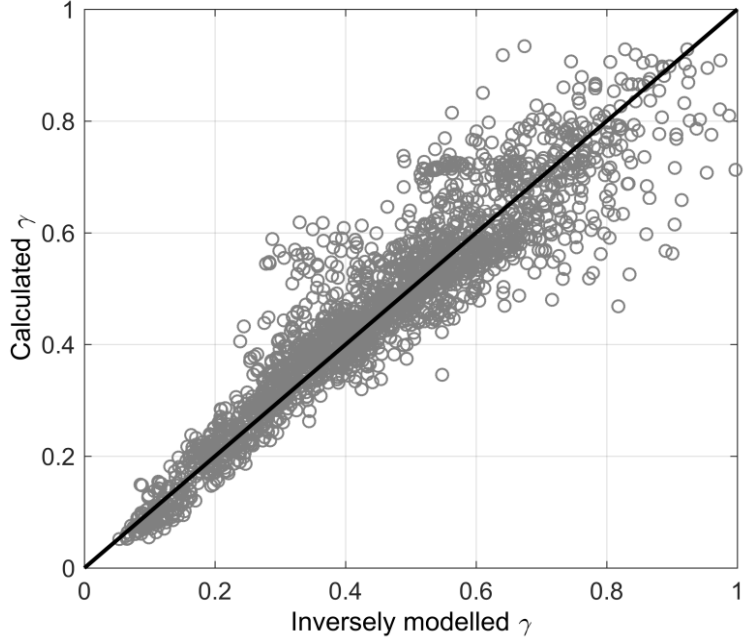

**Fig. 5.** The inversely modelled $\gamma$ versus the calculated $\gamma$ with Eq. (5) with the correlation coefficients $R^2 = 0.86$.

## 5. Model verification

Hereafter we will label the present model as ZL2020 for simplicity. The model performance in predicting $H_{rms}$ using the new breaker index formula (Eq. 5) together with the governing equations Eqs. (1), (2) and (3) is quantitatively evaluated with the root-mean-square percentage error (RMSPE) and the Brier Skill Score (BSS) (e.g., Apotsos et al., 2008; Murphy and Epstein, 1989). RMSPE represents the root-mean-square percentage error with regard to each data record. BSS is used to estimate the percentage error reduction provided by the present model relative to the previous ones. They are expressed as

$$\text{RMSPE} = \sqrt{\frac{1}{N}\sum_{i=1}^{N}(\frac{M_i - O_i}{O_i})^2} \times 100\% \tag{6}$$

$$\text{BSS} = \left[1 - \frac{\text{RMSPE (the present model)}}{\text{RMSPE (the previous models)}}\right] \times 100\% \tag{7}$$

where $N$ is the total number of cross-shore observations, $i$ is the spatial index of data point, $M$ and $O$ are the modeled and observed $H_{rms}$, respectively.

## 5.1. Comparison with R2003 model

Because the present study is conducted on the basis of Ruessink et al. (2003), it is necessary as a first step to compare the performance of the present model (ZL2020) with that of Ruessink et al. (2003) (R2003). Recalling that the governing equations are the same for ZL2020 and R2003, the difference in results should be due to the implementation of different $\gamma$ formulas. Figure 6 shows the measured and predicted wave heights with ZL2020 and R2003 at Duck, Egmond and Terschelling. Both models perform rather well at all coasts, while ZL2020 provides closer prediction to the observation. This is reflected by the median RMSPE, which is 7%, 13% and 12% for ZL2020, and 9%, 15% and 17% for R2003, respectively.

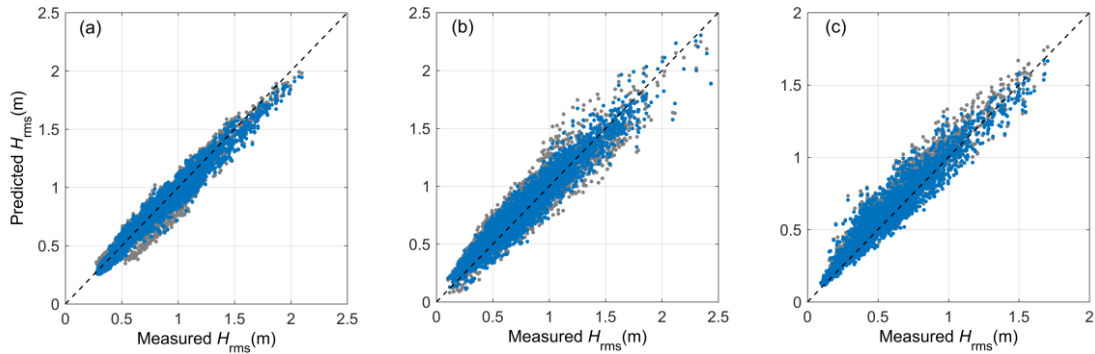

**Fig. 6.** The measured and predicted wave heights with ZL2020 model (blue points) and R2003 model (grey points) at (a) Duck, (b) Egmond and (c) Terschelling.

The most important difference between the $\gamma$ formulas of ZL2020 and R2003 is that ZL2020 considers the possible influence of the offshore wave steepness. To further investigate how the error reduction provided by the new $\gamma$ formula (Eq. 5) depends on $s_0$, Figure 7 presents the time series of $s_0$ and BSS with respect to ZL2020 and R2003 at Duck. BSS > 0 (BSS < 0) indicates that ZL2020 gives lower (higher) errors than R2003. It can be seen that the remarkable error reduction (i.e., BSS > 0) provided by ZL2020 is mostly associated with the situations when $s_0$ is relatively small (e.g., time < 450 h and > 600 h). For relatively large $s_0$ (e.g., time = 450-600 h), both models' performances are comparable but R2003 has a greater accuracy when $s_0$ is highest (e.g., time = 550-600 h). This can be explained by the fact that R2003 was calibrated mainly with the data under wave conditions with larger offshore wave steepness, while ZL2020 was calibrated with the data under all wave conditions. In general, ZL2020 is considered more suitable to predict wave transformation with respect to changing wave conditions, especially for the wave prediction dedicated to simulating onshore sediment transport and beach recovery under mild wave forcing (e.g., Zheng et al., 2014; Eichentopf et al., 2019).

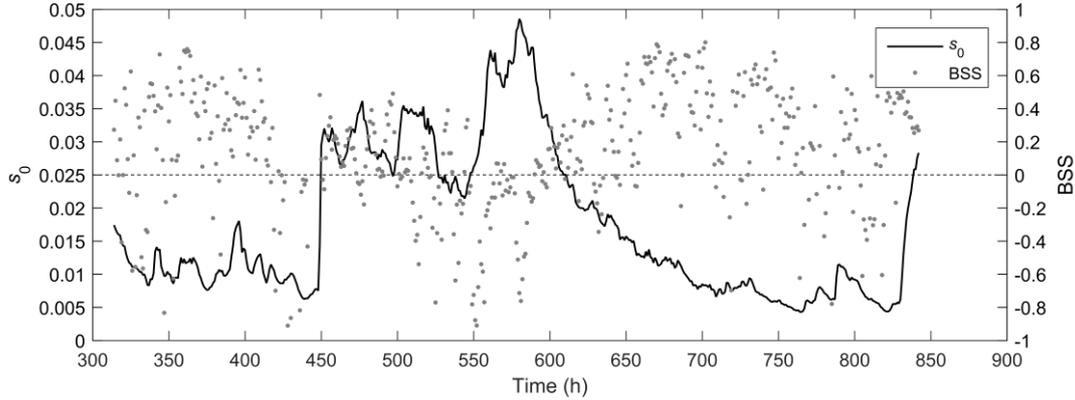

**Fig. 7.** Time series of $s_0$ and BSS with respect to ZL2020 and R2003 at Duck. Time = 0 corresponds to September 21, 1994, 12:00 EST.

When the offshore wave steepness is small, the present $\gamma$ formula (Eq. 5) and that of R2003 describe negative and positive dependence on $kh$, respectively. Figure 8 shows how this distinction affects the model behavior by illustrating two examples of the cross-shore wave distribution at Duck ($t = 735$ h, $s_0 = 0.007$) and Egmond ($t = 553$ h, $s_0 = 0.014$). The inversely modelled and the calculated $\gamma$ with Eq. (5) are in good agreements and both depend negatively on the local water depth (Figure 8a-8b). This favors accurate predictions of the percentage of total (spatially accumulated) energy dissipation ($1-F_w/F_{w0}$, where $F_w$ and $F_{w0}$ are the local and offshore wave energy flux) (Figure 8c-8d), and of the wave height distribution (Figure 8e-8f). However, $\gamma$ calculated with R2003 formula depends positively on the local water depth, producing too large $\gamma$ at most locations (Figure 8a-8b). The consequence of an overpredicted $\gamma$ is the underestimated energy dissipation (Figure 8c-8d) and, as the error accumulates from offshore to shallow water, the systematically overpredicted wave height across the profile (Figure 8e-8f).

Note that three data points at distance = 404.7 m, 564.7 m and 679.6 m existing in Figure 8c and Figure 8e are not shown in Figure 8a. This is because the inversely calculated energy dissipation rates are less than zero (i.e., $D_b < 0$ N/ms) at these locations, and hence cannot be used to derive $\gamma$. In Figure 8c-8d, both the measured data and our model results show noticeable energy dissipation from deep water to shallow water even when the root-mean-square wave height is relatively small and the water depth is relatively deep. Note that, however, the results in Figure 8c-8d are the percentages of total energy dissipation as spatially accumulated over distances of 800 m at Duck and 5000 m at Egmond. A large percentage value at a given location can be due to the small offshore wave energy flux and the long distance of wave propagation, but not necessarily represents an intense local wave breaking. In fact, the local wave energy dissipation rates ($D_b$) are generally lower than 10~20 N/ms for these small wave cases, which are significantly weaker than the values (up to 150 N/ms at Duck and 400 N/ms at Egmond) under large wave conditions.

The results presented in this section demonstrate that the $\gamma$-$kh$ relationship is not invariable and can significantly depend on the offshore wave conditions represented by $s_0$. This should be taken into account in nearshore wave modelling for a more reliable prediction of long-term morphodynamic processes characterized by the time-varying wave climates.

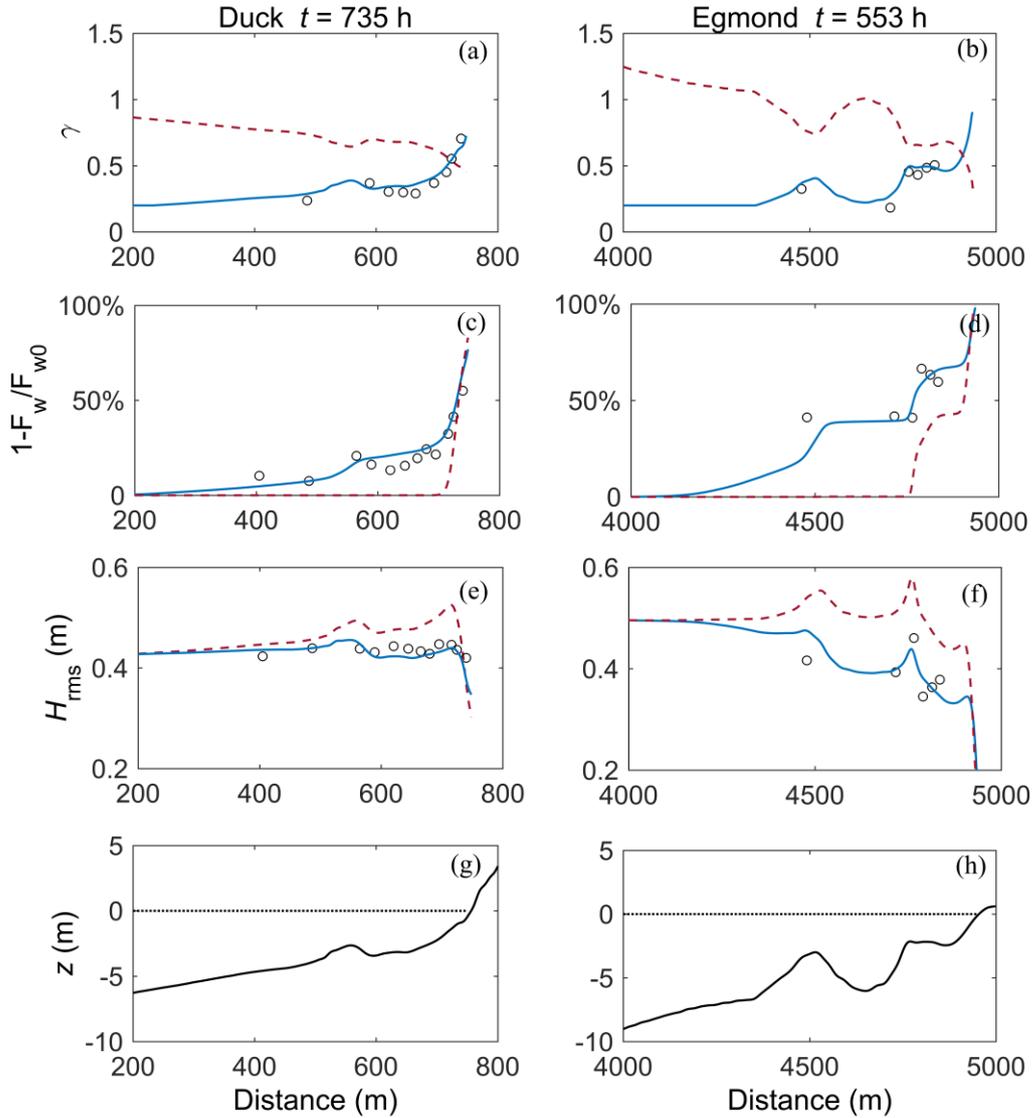

**Fig. 8.** Cross-shore distribution of (a)-(b) the wave breaker index, (c)-(d) the percentage of total energy dissipation, (e)-(f) the root-mean square wave height, (g)-(h) the bed elevation at Duck (left panels) and Egmond (right panels) under wave conditions of small offshore wave steepness, with the measured or the inversely modelled data (circles), the predicted results with ZL2020 model (thick lines) and R2003 model (dashed lines). Distance is relative to the offshore measurement location.

## 5.2. Comparison with other models: Field tests

Using the field datasets in three coasts, the present model (ZL2020) is further compared with seven previous models (R2003, TG1983, JB2007, B1998, BJ1978, S2015, L2017) which are widely used in parametric wave modelling. Since Apotsos et al. (2008) have significantly improved the accuracy of TG1983, JB2007, B1998 and BJ1978 by tuning $\gamma$ for each model based on the similar datasets, here we will employ the newly tuned $\gamma$ formulas of Apotsos et al. (2008) in these models (hereafter TG1983_t, JB2007_t, B1998_t, BJ1978_t) for a more rigorous inter-comparison.

Figure 9 compares the median percentage errors of eight models. All models have the errors between 5% and 20%, consistent with the results of Apotsos et al. (2008). The model error is generally lower at Duck than at two other sites, possibly because most of the models (except S2015 and L2017) have been calibrated with part of (or full) Duck data. ZL2020 shows lowest errors at Duck and Egmond, and very slightly higher error than TG1983_t but still remarkably lower than other models at Terschelling. The BSS values in Table 1 indicated that using ZL2020 model reduces the errors averagely by 16% at Duck, 19% at Egmond, 21% at Terschelling, and by 22% over R2003, 10% over TG1983_t, 15% over JB2007_t, 17% over B1998_t, 20% over BJ1978_t, 23% over S2015, 24% over L2017.

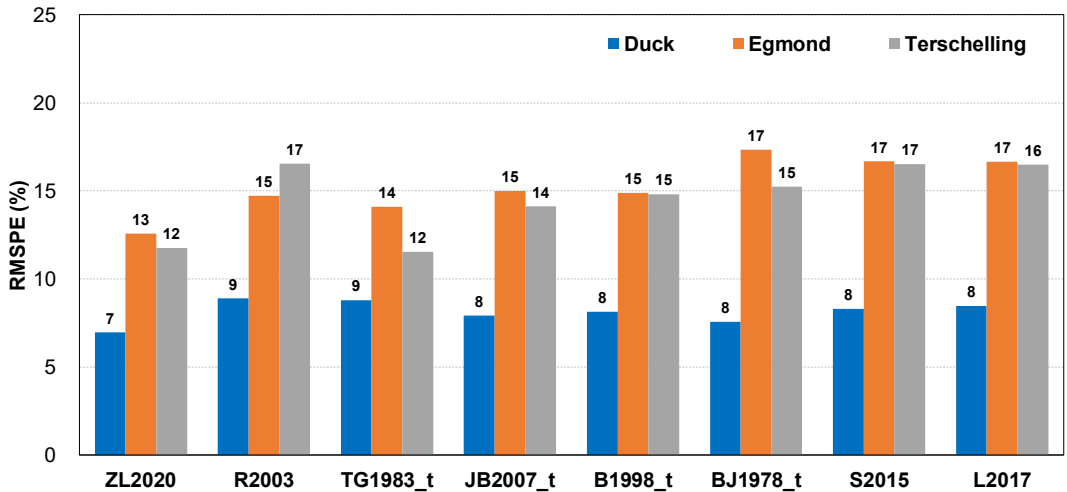

**Fig. 9.** Median percentage errors of eight models in field tests.

Table 1 BSS values (%) using ZL2020 instead of other models

|  | R2003 | TG1983_t | JB2007_t | B1998_t | BJ1978_t | S2015 | L2017 | mean |
|---|---|---|---|---|---|---|---|---|
| Duck | 22 | 21 | 12 | 14 | 8 | 16 | 18 | 16 |
| Egmond | 15 | 11 | 16 | 16 | 28 | 25 | 25 | 19 |
| Terschelling | 29 | -2 | 17 | 21 | 23 | 29 | 29 | 21 |
| mean | 22 | 10 | 15 | 17 | 20 | 23 | 24 |  |

To explore further insights on whether the model accuracy improvement might be related to the offshore wave parameters, Figure 10 presents an example of the median percentage errors of ZL2020 versus $s_0$ at Duck. In this case the ZL2020 model error does not show remarkable dependence on $s_0$. Figure 11 presents the median percentage errors of ZL2020, R2003, TG1983_t and JB2007_t versus $s_0$ at all three coasts. The bin-averaged results shown in Figure 11 may be subject to uncertainties due to different numbers of data points distributed in different $s_0$ bin ranges and to different numbers of observation locations in different coasts. For example, the Duck data was measured at fourteen locations with the minimum water depth $h < 1$ m and $s_0$ ranged from 0.005 to 0.05, but the Terschelling data was measured only at five locations with $h > 2$ m and mostly $s_0 < 0.025$. Nevertheless, it is worthwhile to roughly examine the general model behaviors in Figure 11. For smaller $s_0$ (notably when $s_0 < 0.02$), the significantly highest model accuracy is provided by ZL2020 compared to all other models. Among R2003, TG1983_t and JB2007_t with smaller $s_0$, they have identical accuracy at Duck, but at Egmond and Terschelling the accuracy is TG1983_t > JB2007_t > R2003. For larger $s_0$ (e.g., when $s_0 > 0.02$), the accuracy of ZL2020 is comparable to JB2007_t and R2003 at Duck and Terschelling, while at Egmond R2003 has the highest accuracy than other models. For larger $s_0$, TG1983_t is most accurate at Terschelling but less accurate than other models at Duck and Egmond. In summary, no model provides the best prediction for all $s_0$ at all sites. Nevertheless, the overall best performance is obtained with ZL2020 by significantly reducing the errors under wave conditions with small offshore wave steepness, while not deteriorating its applicability under wave conditions with large offshore wave steepness.

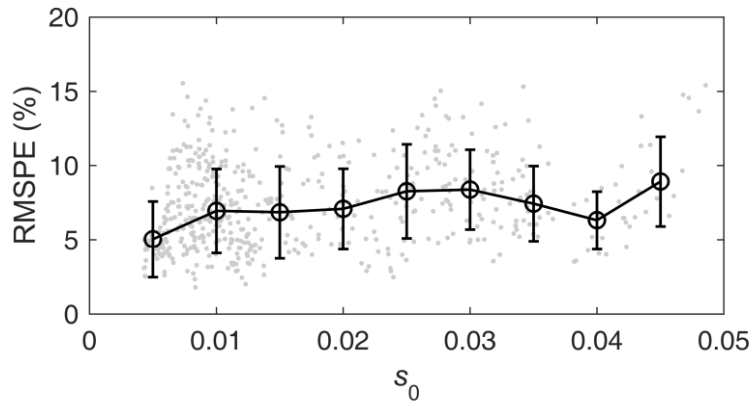

**Fig. 10.** Median percentage errors of ZL2020 versus the offshore wave steepness at Duck. Data are binned corresponding to $s_0 \pm 0.0025$. The mean and standard deviation of each bin range (black circles and lines) and individual data (grey points) are both shown.

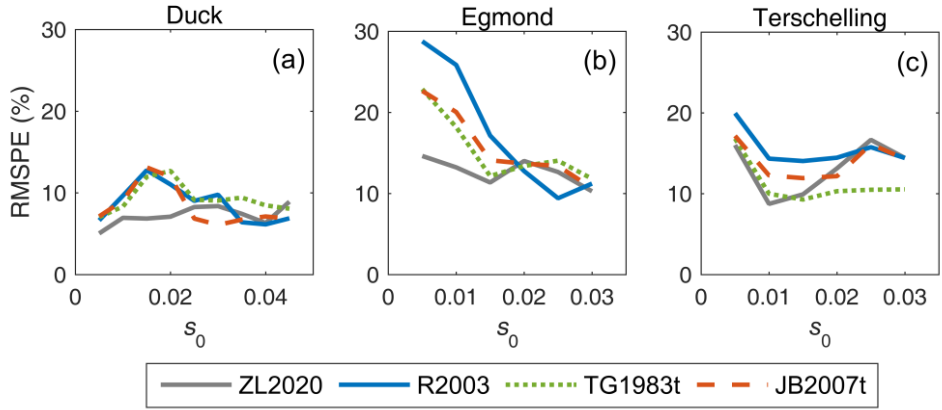

**Fig. 11.** Median percentage errors of ZL2020, R2003, TG1983_t and JB2007_t versus the offshore wave steepness at Duck (a), Egmond (b) and Terschelling (c). Data are binned corresponding to $s_0$ ± 0.0025. Only the bin-averaged results are shown.

## 5.3. Comparison with other models: Laboratory tests

Although the present model (ZL2020) was developed on the basis of field data, it is worthwhile to examine its performance in laboratory tests. In this section, the models are additionally run for 14 laboratory tests. Table 2 provides a summary of the experimental conditions, including wave transformation over sloping and barred beaches with various bed slopes (1:40~1:10), offshore root-mean-square wave heights (0.05~0.43 m), peak wave periods (1.0~4.0 s) in both small- and large-scale wave flumes.

Figure 12 shows the laboratory measured and predicted wave heights with ZL2020, R2003, TG1983_t, JB2007_t, B1998_t, BJ1978_t, S2015 and L2017. It is found that ZL2020, R2003, S2015 and L2017 agree reasonably well with the measured $H_{rms}$, while TG1983_t, JB2007_t, B1998_t and BJ1978_t show an overall underestimation. No remarkable difference in model behavior can be found for wave prediction in small- and large-scale wave flumes that also coincide with sloping and barred profiles, respectively. In general, ZL2020 performs best for the laboratory data used here with a similar accuracy to the field tests. This indicates that although ZL2020 was originally derived from the data collected in field, it is also applicable to reproduce the wave height observed in laboratory.

Table 2 Laboratory experiments used for model verification

| No. | Source | Case | Beach type | Bed slope | $H_{rms0}$ (m) | $T_p$ (s) | Apparatus |
|-----|--------|------|-----------|-----------|---------|--------|-----------|
| 1 | Baldock and Huntley (2002) | J6010A | sloping | 1:10 | 0.10 | 1.67 | small-scale |
| 2 | Baldock and Huntley (2002) | J6010B | sloping | 1:10 | 0.08 | 1.67 | small-scale |
| 3 | Baldock and Huntley (2002) | J6010C | sloping | 1:10 | 0.05 | 1.67 | small-scale |
| 4 | Baldock and Huntley (2002) | J6033A | sloping | 1:10 | 0.10 | 1.67 | small-scale |
| 5 | Baldock and Huntley (2002) | J6033B | sloping | 1:10 | 0.08 | 1.67 | small-scale |
| 6 | Baldock and Huntley (2002) | J6033C | sloping | 1:10 | 0.05 | 1.67 | small-scale |
| 7 | Baldock et al. (1998) | J1 | sloping | 1:20-1:10 | 0.09 | 1.50 | small-scale |
| 8 | Baldock et al. (1998) | J2 | sloping | 1:20-1:10 | 0.07 | 1.50 | small-scale |
| 9 | Baldock et al. (1998) | J3 | sloping | 1:20-1:10 | 0.05 | 1.00 | small-scale |
| 10 | Stive (1985) | S15 | sloping | 1:40 | 0.14 | 1.58 | small-scale |
| 11 | Stive (1985) | S29 | sloping | 1:40 | 0.14 | 2.93 | small-scale |
| 12 | Flores et al. (2016) | run25 | barred | 1:24 | 0.43 | 4.00 | large-scale |
| 13 | Flores et al. (2016) | run27 | barred | 1:24 | 0.37 | 2.70 | large-scale |
| 14 | Scott et al. (2005) | random wave | barred | 1:24 | 0.42 | 4.00 | large-scale |

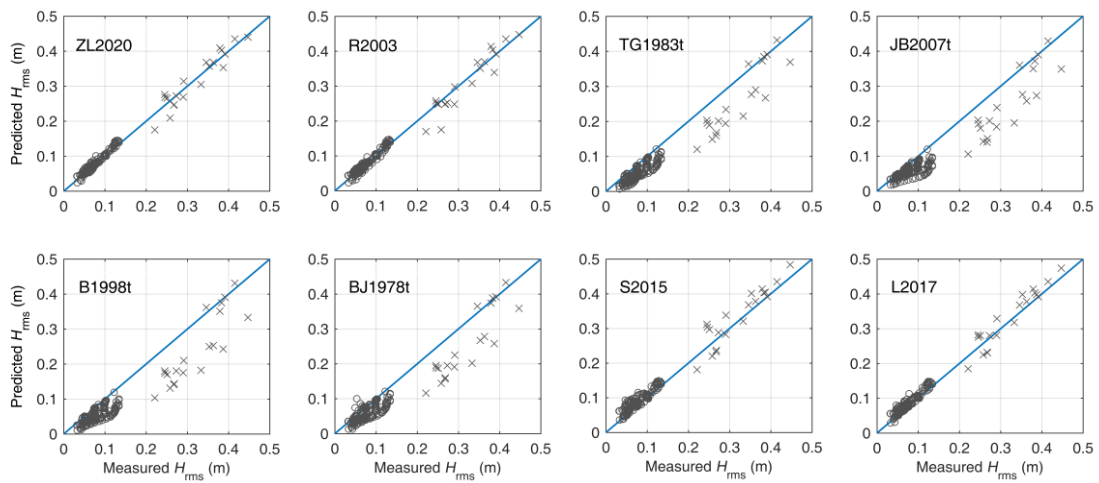

**Fig. 12.** The measured and predicted wave heights in small-scale (○) and large-scale (×) wave flumes with ZL2020 (RMSPE=8%), R2003 (RMSPE=9%), TG1983_t (RMSPE=30%), JB2007_t (RMSPE=25%), B1998_t (RMSPE=30%), BJ1978_t (RMSPE=25%), S2015 (RMSPE=22%) and L2017 (RMSPE=15%).

## 6. Discussion

### 6.1. General behaviors of the new γ formula

The most important feature of the new $\gamma$ formula (Eq. 5) is that it considers the composite dependence of $\gamma$ on both $s_0$ and $kh$ (Figure 3). Figure 13 illustrates the typical $\gamma$-$kh$ relationships provided by Eq. (5). It is interesting to find that the new formula can quantitatively or qualitatively reproduce the results of several previous formulas with respect to different $s_0$ conditions. For a larger $s_0$ (e.g., $s_0 = 0.045$), Eq. (5) well reproduces the results of R2003 formula, and $\gamma$ increases with increasing $kh$ in line with some empirical $\gamma$ parameterizations (e.g., van der Westhuysen, 2010; Salmon et al., 2015; Lin and Sheng, 2017). For a median $s_0$ (e.g., $s_0 = 0.025$), Eq. (5) gives $\gamma$ values with little dependence on $kh$, which is very close to the mean value (0.73) of the calibrated $\gamma$ (with mean $s_0 = 0.022$) in Battjes and Stive (1985) used also as the default value in SWAN model (Booij et al., 1999). For a smaller $s_0$ (e.g., $s_0 = 0.015$), Eq. (5) provides a decreasing $\gamma$ with increasing $kh$. Although not necessarily comparable, such negative $\gamma$-$kh$ relationship appears to be qualitatively consistent with the field and laboratory observed depth-dependence of $H_{rms}/h$, $H_s/h$ or $H_b/h_b$ (e.g., Raubenheimer et al., 1996; Sénéchal et al., 2001, 2005; Power et al., 2010; Rattanapitikon and Shibayama, 2000; Goda, 2010; Robertson et al., 2013, 2015b).

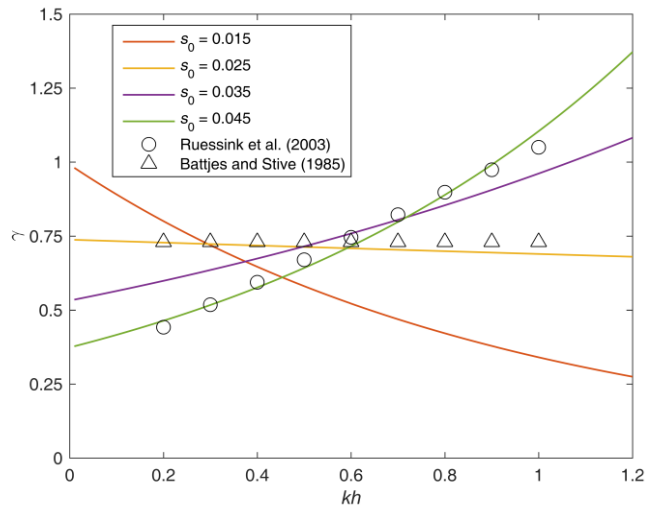

**Fig. 13.** $\gamma$-$kh$ relationships provided by Eq. (5) with respect to different offshore wave steepness (lines), by the formula of Ruessink et al. (2003) (circles), and the mean value (0.73) of the calibrated $\gamma$ (with mean $s_0 = 0.022$) in Battjes and Stive (1985) (triangles) used also as the default value in SWAN model (Booij et al., 1999).

### 6.2. Physical mechanisms for the composite relationships

In this section, we attempt to provide a possible physical interpretation for the composite relationships. First, we consider the normalized water depth ($kh$) as a proxy for local wave nonlinearity. The Ursell number (Ur = $HL^2/h^3$) is widely used to represent wave nonlinearity, which can also be expressed as Ur = $4\pi^2(H/h)/(kh)^2$. Therefore a smaller (larger) $kh$ is corresponding to a larger (smaller) Ur, i.e., a larger (smaller) wave nonlinearity. Similar arguments were also provided in van der Westhuysen (2010).

Second, we suggest two counteractive mechanisms for wave nonlinearity effects on breaking incipience and energy dissipation, namely the breaking intensification (BI) mechanism and the breaking resistance (BR) mechanism. For the BI mechanism, a larger wave nonlinearity (associated with wave breaking at a smaller $kh$) indicates a more asymmetric wave shape with a steeper wave front (Zhang et al., 2017), and a stronger energy dissipation if the wave breaks. In this case, a smaller $\gamma$ is required to be implemented in the models to represent the intensified dissipation (van der Westhuysen, 2010). For the BR mechanism, a larger wave nonlinearity implies that the wave shape has adapted more sufficiently to the local bathymetry due to shoaling, and the wave will struggle not to break until a larger $\gamma$ is achieved, which is the case that a plunging breaker of larger nonlinearity has a larger $\gamma$ than a spilling breaker. In summary, the BI mechanism supports a positive $\gamma$-$kh$ relationship and the BR mechanism supports a negative $\gamma$-$kh$ relationship.

Third, we argue that the offshore wave steepness ($s_0$) affects the relative dominance of BI and BR mechanisms. For larger $s_0$, the smaller surf similarity parameter (Battjes, 1974) favors a saturated surf condition (Power et al., 2010). In this case, significant wave breaking persists from a deeper water depth to the shoreline with a wider surf zone, while the shoaling effect is relatively limited. This implies the BI mechanism dominates with a positive $\gamma$-$kh$ relationship. For smaller $s_0$, the larger surf similarity parameter (Battjes, 1974) favors an unsaturated surf condition (Power et al., 2010). In this case, waves have a longer distance/time to shoal and wave shapes are allowed to deform more sufficiently to adapt to the bathymetry, while intense breaking more likely concentrates in a narrower surf zone of shallower water depth. This implies the BR mechanism dominates with a negative $\gamma$-$kh$ relationship. The above discussion may provide a generic explanation for the opposite $\gamma$-$kh$ relationships affected by $s_0$.

### 6.3. Relevance of wave reflection and bed slope

Wave reflection is one of the factors affecting wave characteristics in the nearshore, especially when the incident wave steepness is small. This process is not normally included in parametric wave models. On a reflective beach where most of the energy is at incident wave frequency (Wright and Short, 1984), random wave reflection is considered to increase the wave height and the height-to-depth ratio mainly close to the shoreline (Martins et al., 2017), so the modelled wave height would be smaller than the measurement. However, our analysis indicates that the previous model systematically overpredicts the wave height across a large portion of the beach profile (Figure 8e-8f). Therefore, wave reflection effect is considered not large enough to explain the overall model-data discrepancy when the offshore wave steepness is small.

In addition, wave reflection is not thought as the main mechanism for the revealed negative $\gamma$-$kh$ relationship (i.e., $\gamma$ tends to increase with decreasing water depth towards the shoreline for smaller $s_0$, shown in Figure 8a-8b), since the similar tendency of $H_s/h$ was observed in Martins et al. (2017) no matter the reflected wave component was included or not.

Consistent with Ruessink et al. (2003) and Power et al. (2010), we do not find a notable correlation between $\gamma$ and the local bed slope, which seems different from some field observations showing a positive dependence of $H_{rms}/h$ on the bed slope (e.g., Raubenheimer et al., 1996; Sallenger and Holman, 1985) or a few parametric wave models using larger $\gamma$ for a steeper slope (e.g., Lin and Sheng, 2017). The reason is yet not very clear. It is noted that the importance of bed slope might be linked to the degree of saturation of surf zone. In contrast to the saturated surf conditions (Raubenheimer et al., 1996), the field data of Power et al. (2010) collected during unsaturated surf conditions shows that $H_{rms}/h$ is controlled by both the local water depth and offshore wave parameters, but has no dependence on the bed slope. Moreover, the present study uses the measured wave heights to derive exact values of $\gamma$ via an inverse modelling approach as Ruessink et al. (2003), while other parametric model studies tuned $\gamma$ to obtain a best-fit wave height prediction.

### 6.4. Wave height data in field experiments

The field datasets of spectral wave height used for the present model calibration and verification, were reconstructed from the pressure measurement with the classic transfer function method (TFM) based on the linear wave theory. Such linear reconstruction method may affect the accuracy of converted wave parameters and therefore the derived breaker index.

Focusing on the unbroken or near-breaking waves and based on the wave-by-wave analysis at the individual wave scale, recent studies have demonstrated that TFM can underestimate the individual wave height or wave shape (e.g., highest crest elevation and surface skewness) by up to 20~30% (e.g., Oliveras et al., 2012; Martins et al., 2017; Bonneton and Lannes, 2017; Bonneton et al., 2018; Mouragues et al., 2019). In this regard, novel fully- and weakly-dispersive nonlinear reconstruction methods were developed by Bonneton and Lannes (2017) and Bonneton et al. (2018), showing robust capability to recover the peaked and skewed shape of the highest waves within wave groups. Whereas for the bulk wave parameters in a random wave train (i.e., wave energy as the focus of the present study), it is realized that TFM gives a reasonable estimate of the spectral wave height (e.g., Bishop and Donelan, 1987; Tsai et al., 2005), with errors generally less than 7.4% even when waves are highly nonlinear in near-breaking conditions (Mouragues et al., 2019), although the energy in the highest frequencies cannot be adequately recovered.

With respect to the breaking waves of asymmetric shape, the application of TFM would be, in principle, not suitable because the linear wave theory does not apply here.

Unfortunately, little knowledge is available to clarify the exact accuracy/error of TFM-based spectral wave height for this case, and no formulation has been specifically developed to handle the proper conversion from pressure to surface elevation under breaking conditions. Nevertheless, some relevant information can be found in a few literatures. For example, Brodie et al. (2015) reported a good agreement between lidar- and pressure-based estimates of significant wave height (approximately 0~1 m) in the inner surf zone with the root-mean-square difference of 0.03~0.07 m, based on the linear reconstruction method with buried pressure sensors. On the other hand, Mouragues et al. (2019) discussed that the hydrostatic reconstruction seems to be the most suitable method here, in view of the facts that the nonlinear shallow-water equations accurately predict wave distortion in the inner surf zone (Bonneton, 2007) where the pressure distribution is assumed mainly hydrostatic (Lin and Liu, 1998; Sénéchal et al., 2001). Indeed, more accurate and direct measurements of the breaking wave surface elevation are still required before a solid conclusion can be drawn on this subject.

For the field wave height data at three sites used in the present study, the TFM was consistently used to convert pressure to water surface elevation. The pressure response factor $K_p = \cosh(kh)/\cosh(k\delta_m)$ where $\delta_m$ is the distance of the pressure sensor from the bottom, was used with a high cutoff frequency ($f_c$) of 0.32 or 0.33 Hz (e.g., Long, 1996; Ruessink et al., 1998), an approach similar to the TFM with sharp cutoff as described in great details in Mouragues et al. (2019). According to the field test at a single near-breaking location of Mouragues et al. (2019), TFM with sharp cutoff at the same $f_c$ (0.32 Hz) could underestimate the spectral wave height by 7% due to the information loss in the highest frequencies. As yet, unfortunately, it remains unclear how this error might vary from offshore to very shallow water including surf zone, which inhibits a definite understanding on the overall accuracy of reconstructed wave heights across the beach profile. Nevertheless, in order to shed more insight into its potential effects on the calibrated wave breaker index ($\gamma$), our preliminary sensibility tests have shown that the calibrated values of $\gamma$ generally increase (decrease) by 7.6% (7.5%) if the reconstructed wave heights at all locations increase (decrease) overall by 7%, but the previously-revealed composite relationship of $\gamma$ to $s_0$ and $kh$ remains unchanged. Noted that this quantity is probably conservative since the wave height error 7% is corresponding to the most nonlinear near-breaking waves, i.e., less error might be expected in other cross-shore regions. If this is the case, the actual difference in $\gamma$ would be less than 7.6% in overall. It is therefore assumed that using the characteristic wave height converted from pressure may not significantly affect the main findings in the present study. This is also supported by the good agreements between the modelled wave height and both field and laboratory data with the similar accuracies. We finally note that a proper reconstruction method is indeed required if concerns are laid on the individual nonlinear wave parameters or the phase-resolving wave model validation (Bonneton et al., 2018). Further investigations on the reconstruction method suitable for breaking waves are also needed.

## 7. Conclusions

Using the field datasets of wave height in three coasts and the inverse modelling approach, we revisit the parameterization of wave breaker index ($\gamma$) employed in the parametric nearshore random wave transformation models. A composite dependence of $\gamma$ on both the offshore wave steepness ($s_0$) and the normalized local water depth ($kh$) is revealed that, $\gamma$ increases with increasing $kh$ for larger $s_0$, and $\gamma$ decreases with increasing $kh$ for smaller $s_0$. The opposite $\gamma$-$kh$ relationships within different ranges of $s_0$ can be explained by the wave nonlinearity effects via two counteractive physical mechanisms, i.e., the breaking intensification mechanism and the breaking resistance mechanism. This provides more insights into the wave breaking behavior, compared to the previous understanding of the positive $\gamma$-$kh$ relationship independent of $s_0$, or the positive $\gamma$-$s_0$ relationship independent of $kh$.

A new $\gamma$ formula (Eq. 5) is proposed to account for such composite relationships. Implementation of this new $\gamma$ formula in a parametric wave model systematically reduces the median percentage error of wave height prediction by 10~24% (mean = 19%) relative to seven widely used models in literatures. The error reduction arises mainly from the improved prediction under wave conditions with small offshore wave steepness, which is important for studying onshore sediment transport and beach recovery. The present study is anticipated to promote more reliable prediction of long-term coastal morphodynamic processes involving time-varying wave climates.

## Acknowledgments

The authors are very much grateful to B.G. Ruessink for kindly providing the field datasets and valuable information. This work was supported by the National Natural Science Foundation of China (51879096), the Key Program of National Natural Science Foundation of China (41930538), the National Science Fund for Distinguished Young Scholars (51425901), and the Special Research Funding of State Key Laboratory of Hydrology-Water Resources and Hydraulic Engineering (20195025812, 20185044512).